\documentclass[11pt]{article}
\catcode`\@=11
\def\marginnote#1{}

\newcount\hour
\newcount\minute
\newtoks\amorpm
\hour=\time\divide\hour by60
\minute=\time{\multiply\hour by60 \global\advance\minute by-\hour}
\edef\standardtime{{\ifnum\hour<12 \global\amorpm={am}%
        \else\global\amorpm={pm}\advance\hour by-12 \fi
        \ifnum\hour=0 \hour=12 \fi
        \number\hour:\ifnum\minute<10 0\fi\number\minute\the\amorpm}}
\edef\militarytime{\number\hour:\ifnum\minute<10 0\fi\number\minute}

\def\draftlabel#1{{\@bsphack\if@filesw {\let\thepage\relax
      \xdef\@gtempa{\write\@auxout{\string
          \newlabel{#1}{{\@currentlabel}{\thepage}}}}}\@gtempa \if@nobreak
    \ifvmode\nobreak\fi\fi\fi\@esphack} \gdef\@eqnlabel{#1}}
    \def\@eqnlabel{}
\def\@vacuum{}
\def\draftmarginnote#1{\marginpar{\raggedright\scriptsize\tt#1}}

\def\draft{
  \oddsidemargin -.5truein
  \def\@oddfoot{\footnotesize \sl preliminary draft \hfil
    \rm\thepage\hfil\sl\today\quad\militarytime}
  \let\@evenfoot\@oddfoot \overfullrule 3pt
    \let\label=\draftlabel
    \let\marginnote=\draftmarginnote
  \def\@eqnnum{(\theequation)\rlap{\kern\marginparsep\tt\@eqnlabel}%
    \global\let\@eqnlabel\@vacuum}

  }

\voffset= - 1.2in
\hoffset= - 1.0in

\def\be{\begin{equation}}
\def\ee{\end{equation}}
\def\bea{\begin{eqnarray}}
\def\eea{\end{eqnarray}}
\def\be{\ba}
\def\ee{\ea}
\def\<{\langle}
\def\>{\rangle}
\def\stackreb#1#2{\mathrel{\mathop{#2}\limits_{#1}}}

\def\d{\partial}
\def\N2{${\cal N}=2$}

\def\tr{{\mathrm{tr\,}}}

\def\1N{${\cal N}=1$}
\def\4N{${\cal N}=4$}

\def\bea{\begin{eqnarray}}
\def\eea{\end{eqnarray}}
\def\bqa{\begin{eqnarray}}
\def\eqa{\end{eqnarray}}

\def\beq{\begin{equation}}
\def\eeq{\end{equation}}
\def\ba{\beq\begin{array}{c}}
\def\ea{\end{array}\eeq}
\def\be{\beq}
\def\ee{\eeq}

\let\text=\mathrm

\def\beq{\begin{equation}}
\def\eeq{\end{equation}}
\def\bea{\begin{eqnarray}}
\def\eea{\end{eqnarray}}

\renewcommand{\d}{{{\partial}}}

\renewcommand{\<}{\langle}
\renewcommand{\>}{\rangle}

\def\2{{1\over 2}}
\def\stackreb#1#2{\mathrel{\mathop{#2}\limits_{#1}}}

\def\Tr{{\tr}}

\def\d{\partial}

\def\â{$\tau$}

\newcommand{\cpict}[3]{
\dimen1=#1\advance\dimen1 by-\hsize\divide\dimen1 by-2
\vtop to #2{
\noindent\hskip\dimen1{\special{em:graph #3.bmp}}
\vfil}\hskip-2cm
}

\let\@@savethanks\thanks
\def\thanks#1{\gdef\thefootnote{\alph{footnote}}\@@savethanks{#1}}

\title{{\bf Matrix Models vs. Matrix Integrals
}
\vspace{.5cm}}
\author{{\bf A. Mironov}\footnote{E-mail:
\ mironov@itep.ru; mironov@lpi.ru}
\date{ } \\
{\small {\it Lebedev Physics Institute}
and {\it ITEP, Moscow, Russia}}
}

\begin{document}

\maketitle

\vspace{-6.5cm}

\begin{center}
\hfill FIAN/TD-10/05\\
\hfill ITEP/TH-45/05\\
\end{center}

\vspace{3.5cm}

\begin{abstract}
In a brief review, we discuss interrelations between arbitrary solutions of
the loop equations that describe Hermitean one-matrix model and particular
(multi-cut) solutions that describe concrete matrix integrals. These latter
ones enjoy a series of specific properties and, in particular, are
described in terms of Seiberg-Witten-Whitham theory. The simplest example of
ordinary integral is considered in detail.
\end{abstract}
\def\thefootnote{\arabic{footnote}}

\section{Introduction}

Recent interest in matrix models and especially in their
multi-cut solutions was inspired by the studies in
${\cal N}=1$ SUSY gauge
theories due to Cachazo, Intrilligator and Vafa \cite{CIV}
and by the proposal of Dijkgraaf and Vafa \cite{DV} to calculate the
low energy superpotentials, using the partition function of multi-cut solutions.
The solutions themselves are well-known already for a long time
(see, e.g., \cite{JU90}) with a new vim due to the paper
by Bonnet, David and Eynard \cite{David}.

The matrix model under consideration is the Hermitean one matrix model. Its
partition function is given by the integral over $N\times N$ Hermitean
matrix
\be\label{mamoint}
Z_N(t)\equiv {1\over\hbox{Vol}_{U(N)}}\int DM \exp\left(
\Tr\sum_k t_kM^k\right)
\ee
Here $DM$ is the invariant (Haar) measure on Hermitean matrices, which is
just the flat measure and $\hbox{Vol}_{U(N)}$ is the volume of the unitary group
$U(N)$ \cite{versus}.
Since the integrand in (\ref{mamoint}) is invariant w.r.t. matrix conjugation,
one can reduce formula (\ref{mamoint}) to an integral over eigenvalues of the
matrix $M$ integrating out the angular variables. It can be done using
formulas from \cite{Mehta}, and the result reads
\be\label{eigen}
Z_N(T)={1\over N!}\int \prod_i dx_i\Delta^2(x)\exp\left(\sum_{k,i}t_kx_i^k
\right)
\ee
where $\Delta (x)$ is the Vandermonde determinant, $\Delta (x)\equiv\det_{i,j}
x_i^{j-1}=\prod_{i>j}(x_j-x_i)$.

The integrals (\ref{mamoint}) and (\ref{eigen})
still need to be defined. Indeed, one has to fix the integration contours in
these multiple integrals. Moreover, these contours can differ from each
other. Saying this, one definitely forgets that we originally started from
Hermitean matrices. However, it is in no way important for the properties
of (\ref{mamoint}) and (\ref{eigen}). In fact, the word ``Hermitean" has no
other sense but fixing the integration contour. What we really need to fix
in (\ref{mamoint}) is the flat measure. Say, this is enough to obtain
(\ref{eigen}) from (\ref{mamoint}).

In order to define (\ref{mamoint}) and (\ref{eigen}) one can substitute them
with their saddle point approximations \cite{David,KMT}.
The other possibility \cite{AMM1,AMM2,AMM3} is to observe that they satisfy
an infinite set of loop (Virasoro) equations
(=Schwinger-Dyson equations, =Ward identities) \cite{loop,dVir}. Indeed, this is
the consequence of only the flatness of measure and invariance of integrals
w.r.t. the change of variables.

The second possibility means that one calls the matrix model partition function
any solution to the loop equations. Then, the partition function is not
a function but
a formal $D$-module, i.e. the entire collection of power series
(in $t$-variables), satisfying a system of consistent linear
equations. Solution to the equations does not need to be unique, however,
an appropriate analytic continuation in $t$-variables transforms
one solution to another, and, on a large enough moduli space
(of {\it coupling constants} $t$), the whole entity can be considered,
at least formally, as a single object: this is what we call
{\it the partition function}.
Naively different solutions are interpreted as different {\it branches}
of the partition function, associated with different {\it phases} of the theory.
Further, solutions to the linear differential
equations can be often represented as integrals (over {\it spectral
varieties}), but integration ``contours'' remain unspecified: they can
be generic {\it chains} with complex coefficients (in the case of integer
coefficients this is often described in terms of {\it monodromies}, but
in the case of partition functions the coefficients are not required
to be integer).

Therefore, in further consideration we distinguish between {\it matrix model},
which is a set of solutions to the loop equations, and {\it matrix integral}
which is an integral with a specified integration contour. Note that the
matrix model partition function is an arbitrary linear combination of a
proper set of basis matrix integrals.

The natural choice for this basis is given by the multi-cut
(or Dijkgraaf-Vafa) solutions mentioned above. They are distinguished by a special
property of isomonodromy that allows one to associate a
Seiberg-Witten-Whitham system with them \cite{CM,CMMV2}, the
corresponding partition function having a
multi-matrix model integral representation \cite{David,KMT}.

\section{$1\times 1$ matrix case: a toy example}

We start with the simplest example of the $1\times 1$ matrix, i.e. the matrix
integral is just an ordinary one-fold integral. Let us start first from the
integral with potential of the very general form and see what can be
done.

\subsection{Loop equations}

Thus, we start from the integral
\be\label{mamo1}
Z(\{t_k\})=\int_C dx \exp\left(\sum_k t_kx^k\right)
\ee
At the moment we do not specify the integration contour $C$, but assume it
is closed, or, at least, the integrand is canceled at the ends of the contour
with any monomial $x^k$ of arbitrary degree. Then, one can immediately obtain
an infinite set of equations\footnote{These equations are called loop
equations, Virasoro constraints, Schwinger-Dyson equations, Ward identities,
see \cite{dVir}.} satisfied by this integral. To obtain it, one suffices
to consider the integral of the full derivative
\be\label{fd}
\int_C dx{\partial\over\partial x}
\left[x^{n+1}\exp\left(\sum_k t_kx^k\right)\right],\ \ \ \ n\ge -1
\ee
which is zero. (\ref{fd}) gives rise to the set of constraints
\be\label{Vir1}
\sum_k kt_k{\partial^{k+n} Z\over\partial t_1^{k+n}} +(n+1){\partial Z\over\partial
t_n}=0,\ \ \ \ n\ge -1
\ee
where ${\d Z\over\d t_0}\equiv Z$.
Note that
\be
{\d Z\over\d t_k}={\d^k Z\over\d t_1^k}
\ee
and, therefore, the first terms of the sum in (\ref{Vir1}) can be rewritten as
the linear derivatives w.r.t. the couplings $t_{k+n}$.

Let us try to solve (\ref{Vir1}) as a power series in $t_k$'s,
\be\label{Taylor}
Z=c^{(0)}+\sum_kc^{(1)}_kt_k+\sum_{k,l}c^{(2)}_{kl}t_kt_l+...
\ee
Then, from the constraint with n=0 we immediately obtain $c^{(0)}=0$,
similarly, $c^{(1)}_k=0$ etc. This means the solution is trivial: $Z=0$. In
fact, this should not come as a surprise, since the dimension of coupling $t_k$
is equal to $k$, and the dimension of $c^{(l)}$ should be negative. However,
we have no quantities with any negative dimensions at hands.

In order to get a non-trivial solution, we need to allow, at least some
couplings $t_k$ to be in the denominator. Let us fix a few (say, $p$)
first couplings not to be small, i.e. we shift these couplings $t_k\to
T_k+t_k$, $k=1,...,p$ and consider $Z$ as a power series in $t_k$'s but not in
$T_k$'s. Then (\ref{Vir1}) has non-trivial solutions.

In other words, we fix a polynomial $V(x)\equiv\sum_k^pT_kx^k$ and consider
the integral in (\ref{mamo1}) as taken perturbatively w.r.t. to $t_k$'s,
\be\label{mamo12}
Z(\{t_k\})=\int_C dx \exp\left(V(x)+\sum_k t_kx^k\right)
\ee
i.e. one calculates the moments
\be\label{moment}
\int_C dx x^k e^{V(x)}
\ee
The generating function (resolvent) for these moments is
\be\label{res}
G(z)\equiv \int_C dx  {e^{V(x)}\over z-x}=\sum_k{1\over z^{k+1}}\int_C dx x^k e^{V(x)}
\ee
where integral is understood as the principal value integral.

\subsection{``Matrix" model}

Below we consider the simple example of the cubic potential $V(x)$. By the
shift of $x$ one can erase the quadratic term in the potential and, by
rescaling $x$, make $T_3$ equal to $-1/3$. Therefore, we are left with
the only essential variable, which we choose at the moment to be $T_1$.
Constraints (\ref{Vir1}) taken at zero couplings reduce then to the single
equation
\be\label{Airy}
{d^2Z\over dT_1^2}+T_1Z=0
\ee
which is nothing but the Airy equation. It has two solutions
\be
Z(T_1)=\int_C dx\exp\left(-{x^3\over 3}-T_1x\right)
\ee
that
correspond to two basically different choices of the integration contour
$C$ in (\ref{mamo12}).
The contour has to be chosen so that the integrand
vanishes at its ends, i.e. the contour should go to infinity where
$\Re\hbox{e}
x^3<0$. One of the possibilities is to choose the imaginary axis as $C$.
This gives the standard Airy function
\be\label{Ai}
\hbox{Ai}(T_1)=\int_0^{\infty}\cos\left({x^3\over 3}-T_1x\right) dx
\ee
Another independent solution to the Airy equation should be associated with
a contour that connects other infinities. Say, one can choose the contour
that goes along the imaginary axis from $+\infty$ to zero and then goes
along the positive ray of the real axis. We call the corresponding
function $\hbox{Ai}_2(T_1)$.\footnote{The difference
$\hbox{Ai}_2(T_1)-\hbox{Ai}(T_1)$ is usually
denoted $\hbox{Bi}(T_1)$, see \cite{AS}, formula (10.4.33).}
Then, the ``matrix" model (general) solution is, in our case,
\be
Z(T_1)=\xi \hbox{Ai}(T_1)+\zeta \hbox{Ai}_2(T_1)
\ee
where $\xi$ and $\zeta$ are arbitrary constants. Now one can use equations
(\ref{Vir1}) to generate, recursively and unambiguously, $Z$ as a power
series in $t_k$'s.

Thus, we can define the ``matrix" model partition function
$Z(\{t_k\})$ as a solution to the defining set of equations,
(\ref{Vir1}). Then, the entire freedom that we have in our
``matrix" model is due to the constants $\xi$, $\zeta$. Choosing these constants,
that is, choosing a formal sum of the distinct integration contours $C$,
fixes the ``matrix" model solution uniquely.

\subsection{``Matrix" integral}

In contrast to the ``matrix" model, the ``matrix" integral is defined by an
integral. The freedom one then has is in choosing the integration contour.
Therefore, one can take a basis in the space of all solutions by choosing
some basis contours. This gives us a basis of ``matrix" integrals, or the
``matrix" integral provided with an index associated with the set of basis
contours. Then, constructing linear combinations of these integrals, one
obtains the general solution to the ``matrix" model.

To perform effective calculations, one has to make a clever choice of the
basis of integration contours. They should be naturally associated with the
asymptotical expansion of the integrals. Say, dealing with cubic potential,
one would better choose the basis contours corresponding to
the solutions of the Airy equation controlled by different quasiclassical
expansions. Indeed, let us make an
asymptotic expansion of (\ref{Ai}) at large $T_1$. Then, the saddle point
equation has the two solutions $x=\pm\sqrt{T_1}$. Depending on the choice of the
integration contour $C$, one should choose one or the other solution and
expand the integral around this solution to obtain the asymptotic expansion.

Generally, there are $p-1$ solution to the saddle point equation $V'(x)=0$,
and exactly that many different quasiclassical expansions and basis contours.

Now, in order to obtain $Z$ as a function of all $t_k$, one may use two
different strategies. First of all, one can just iteratively solve equations
(\ref{Vir1}). The other possibility is to calculate moments (\ref{moment})
using properties of the corresponding ``matrix" integrals, i.e. in the cubic
case, those of the Airy functions.

There is also a more tricky possibility which is in no way technically simpler, but
will be of great use later. That is, let us consider a possibility of
immediate calculating integral (\ref{mamo12}). This integral is typically
not a power series in the higher couplings. Indeed, since the degree of
potential is now much higher than $p$, there are much more possibilities of
choosing the integration contours (=quasiclassical regimes, =solutions to
the saddle point equation). We have to fix the integration contours so that
to have a smooth (power) behaviour upon bringing higher couplings to zero,
while the behaviour of the
integral w.r.t. the first $p$ couplings remains arbitrary. This leaves us with
the freedom of exactly $p$ different integration contours.

Let us see how it works in the cubic case. Here we have two possibilities of
choosing the integration contours. Note that the Airy equation (\ref{Airy})
can be reduced to the Bessel equation and, therefore, its solution is
expressed via the cylindric functions. More concretely, the function
$\sqrt{T_1}Z_{1/3}({2\over 3}T_1^{3/2})$ solves the Airy equation, \cite{AS},
where $Z_{1/3}(z)$ is any cylindric function of order $1/3$.
Now let the two basis cylindric functions be the Hankel functions of the first
and second kinds $H^{(1,2)}(z)$
(this choice corresponds not to $\hbox{Ai}(T_1)$ or
$\hbox{Ai}_2(T_1)$, but to their linear
combinations). We also restore the dependence on all the three couplings
(and slightly rescale them for the sake of convenience). Then,
\be
Z(T_1,T_2,T_3)=\int dx\exp\left(iT_3x^3-T_2x^2+iT_1x\right)=
\sqrt{\eta}Z_{1/3}({2\over 3}\eta^{3/2}),\ \ \ \
\eta\equiv {T_2^2\over (3T_3)^{4/3}}-{T_1\over (3T_3)^{1/3}}
\ee
and, using the asymptotic expansion of the Hankel functions, we
find that the smooth behaviour under $T_3\to 0$, i.e. $\eta\to\infty$
is celebrated with  $H^{(2)}(z)$ (see \cite{BE}, formula (7.13.2)),
\be
H^{(2)}_{1/3}(z)\stackreb{z\to\infty}{\sim}\sqrt{{2\over\pi z}}
\exp\left(-i{4z-2\pi\nu-\pi\over 4}\right)\sum_{m=0}{({1\over 3},m)\over
(2iz)^m},\ \ \ \ ({1\over 3},m)\equiv {\Gamma\left({5\over 6}+m\right)\over
m!\Gamma\left({5\over 6}-m\right)}
\ee
Therefore, we finally have (here we put $T_1=0$ for the sake of simplicity)
\be
Z(T_1,T_2,T_3)={1\over\sqrt{T_2}}\sum_{m=0}(-)^m\left({1\over 3},m\right)
\left({27T_3^2\over 4T_2^3}\right)^m
\ee
i.e., with this choice of solution, $Z$ is indeed a power series in $T_3$.
One can easily check that
the corresponding power coefficients coincide with the
corresponding moments of the
Gaussian integral. Therefore, one may really calculate moments of the Gaussian
integral in this tricky way.

\subsection{Resolvent and loop equation}

To conclude our discussion of the toy example, we comment on the properties of
the resolvent (\ref{res}). We start from the simplest Gaussian potential.
Then, the resolvent
\be
G(z)\equiv\int_{-\infty}^{\infty}dx{e^{-T_2x^2}\over z-x}=
2z\int_{-\infty}^{\infty}dx{e^{-T_2x^2}\over
z^2-x^2}= w(\sqrt{T_2}z),\ \ \ \ w(z)\equiv \int_{-\infty}^{\infty}
dx{e^{-x^2}\over z-x}
\ee
can be calculated in two different ways. First of all,
one can use the formula
\be
{1\over z-x}=\int_0^{\infty}due^{-u(z-x)}
\ee
and further calculating the Gaussian integral to express the resolvent
through the error function\footnote{One should be careful with integrating
around the point $z=x$.},
\be\label{resGi}
w(z)=e^{-z^2}\left(-i\pi+2\sqrt{\pi}\int_0^ze^{x^2}dx\right)= e^{-z^2}
erfc(-iz),\ \ \ \ \ erfc(z)\equiv -i2\sqrt{\pi}\int_z^{\infty}e^{-x^2} dx
\ee
The other possibility is to find the differential equation for $w(z)$. To
this end, let us again use vanishing of the integral of the full derivative,
\be
0=\int_{-\infty}^{\infty}dx{\d\over\d x}\left({e^{-x^2}\over z-x}\right)
\ee
This leads to the equation
\be\label{resdifG}
w'(z)=2\sqrt{\pi}-2zw(z)
\ee
Solving this equation, one can arrive at (\ref{resGi}) again.

In the case of more complicated potentials $V(x)$, one can not manifestly
calculate the integral for $G(z)$. Moreover, the differential equation is
also quite complicated, with order increasing as the degree of the polynomial
$V(x)$ increases.
However, there is a universal form of the equation for the resolvent
that we discuss now. To obtain it, let us consider the next simple example
of the cubic potential. Then, we can write, as above,
\be
0=\int_Cdx{\d\over\d x}\left(
{e^{{x^3\over 3}-T_1x}\over z-x}\right)
\ee
which is equivalent to
\be
G'(z)=V'(z)G(z)-\int_Cdx xe^{{x^3\over 3}-T_1x}
-4z
\int_C dxe^{{x^3\over 3}-T_1x}
\ee
In contrast to the Gaussian integral, one can not manifestly calculate
moments of $e^{{x^3\over 3}-T_1x}$. Moreover, these moments contain an
ambiguity related to the choice of the integration contour.

Note, however, that the uncalculable part is
a (linear) polynomial. Therefore, one obtains that the following equation
is correct
\be\label{resdif}
G'(z)=[V'(z)G(z)]_-
\ee
where $[...]_-$ denotes the projector onto negative powers of $z$, and we took
into account that $[G'(z)]_-=G'(z)$. Looking at (\ref{resdifG}), one
observes that it satisfies (\ref{resdif}) too. Moreover, repeating the
derivation for the general polynomial potential, one comes to the same
universal result (\ref{resdif}).

\section{$N\times N$ matrix case}

Now we consider the true matrix integral following mainly the line discussed
in the previous section.

\subsection{Loop equations}

We start now with integral (\ref{mamoint}), again not specifying
integration contours. (Since this is the multiple integral,
we have freedom to choose different contour for each integration.)
Let us first obtain the set of equations satisfied by this integral. As
before, we consider the integral of the total derivative
\be
\int DM\Tr\left[{\d\over\d M^t}\left(M^{n+1}\exp\left(
\Tr V(M)+\sum_k t_k\Tr M^k\right)\right)\right]=0
\ee
where $M^t$ is the transpond matrix. These equations lead to the
following set of constraints
(=Schwinger-Dyson equations,=Ward identities) \cite{dVir}
\be
\hat L_m Z_N(t) = 0, \ \ m\geq -1
\label{Vir2}\\
\hat L_m = \sum_{k\geq 0} k\left(t_k\right)
\frac{\partial}{\partial t_{k+m}} +
\sum_{\stackrel{a+b=m}{a,b\geq 0}} \frac{\partial^2}{\partial t_a\partial t_b}
\ee
where ${\d Z_N\over\d t_0}\equiv NZ_N$. Note that this time we can not replace
derivatives w.r.t. higher couplings with higher order derivatives w.r.t. the
first coupling. As before, in order to have non-trivial solutions to these
equations, we need to shift first $p$ couplings: $t_k\to T_k+t_k$, $k=1,...,p$
and then deal with the partition function (\ref{mamoint}) as a power series
in the couplings $t_k$'s. Then, one also needs to add to constraints
(\ref{Vir2}) the conditions
\be
{\d Z_V\over\d T_k}={\d Z_V\over\d t_k} \ \ \ \ \forall k=0,...,n+1
\ee

However, since now we have the set of equations w.r.t. various couplings,
they can no longer be reduced to a single equation w.r.t., say, $T_1$.
Therefore, the freedom in solving equations (\ref{Vir2}) is much larger.

\subsection{Matrix model}

How many solutions to equations (\ref{Vir2}) do we expect now? To understand
this, let us again study an oversimplified example. Namely, look at the
model similar to (\ref{eigen}), but without the Vandermonde determinant.
Then, it reduces to the product of independent factors
\be
Z_N(T|t)=\prod_i \int_{C_i}dx_i \exp\left(\sum_k t_kx_i^k\right)
\ee
Each factor here is a solution to the corresponding ordinary
differential equation. Say, in the cubic case, one has
\be
Z_N(T|t)=\prod_i^N\left[\xi_i\hbox{Ai}(\eta)+\zeta_i\hbox{Ai}_2(\eta)\right]
=\sum_{k}\Xi_{k}\left[\hbox{Ai}(\eta)\right]^{k}
\left[\hbox{Ai}_2(\eta)\right]^{N-k}
\ee
where $\Xi_{k}$ are the coefficients constructed from products of
$\xi_i$'s and $\zeta_i$'s which determine the freedom in the matrix model
partition function. Since $\Xi_k$ are arbitrary coefficients, one may
interpret them as counterparts of Fourier series coefficients of an
arbitrary function $Z_N(\eta)$ of $\eta$, the Fourier
exponentials being substituted with the combination of the Airy functions.

Generally, as we discussed in the previous section,
for the polynomial potential of degree $p$, there are $p-1$
independent basis functions, i.e. the freedom in the matrix model partition
function is an arbitrary function of $p-1$ variables. The same counting
certainly remains valid for the model with the Vandermonde determinant
present, (\ref{eigen}).

Let us now understand the origin of the arbitrary function of $p-1$
variables in terms of constraints (\ref{Vir2}). To this end, note that
if one solves them recurrently, expanding the partition function into the
power series in couplings $t_k$'s (like (\ref{Taylor})), only the first two
constraints of (\ref{Vir2}) are really restrictive for $Z_N(T)$,
while all other constraints just allow one to restore recurrently the dependence
on the couplings $t_k$'s. As we already explained, (\ref{Vir2}) are less
restrictive than (\ref{Vir1}) since they can not be reduced to an ordinary
differential equation in $T_1$.

The first two constraints of (\ref{Vir2})
are linear in derivatives and, therefore, we
consistently truncate them to $t=0$ and then express two
derivatives, say, $\partial Z/\partial T_{p+1}$ and
$\partial Z/\partial T_{n}$ through $\partial Z/\partial T_{l}$
with $l = 0,\ldots, p-1$. As a corollary,
the partition function can be represented as
\be
Z_N(T)=
\int dk z(k;\eta_2,\ldots,\eta_p)
e^{\frac{1}{\hbar}(kx-k^2w)}
\label{etaparam}
\ee
with an arbitrary function $z$ of $p$ arguments $(k,\eta_2,\ldots,\eta_p)$.
Here the following variables invariant w.r.t. the first constraint are used,
\be
w = \frac{1}{p+1}\log T_{p+1}, \ \ \
x  = T_0 + \ldots \sim \eta_{p+1}
\ee
\be
\eta_k = \left(T_p^k + \frac{k(k-2)!}{p!}\sum_{l=1}^{k-1} (-)^l
\frac{(p+1)^l (p-l)!}{(k-l-1)!}
T_{p-l}T_p^{k-l-1}T_{p+1}^l
\right)T_{p+1}^{-\frac{kp}{p+1}}
\ee
We discussed a particular case of these formulas in s.2.3.

\subsection{Matrix integral}

Let us discuss now what could be a choice of basis matrix integrals. We
basically repeat the procedure we applied in the ordinary integral case,
i.e. use the saddle point approximation. Different saddle points
$M = M_0$ are given by the
equation $V'(M_0) = 0$. If the polynomial
\be
V'(x) = \prod_{i=1}^p(x-\alpha_i)
\ee
has roots $\alpha_i$, then, since $M_0$ are matrices
defined modulo $U(N)$-conjugations (which allow one
to diagonalize any matrix and permute its eigenvalues),
the different saddle points are represented by
\be
M_0 =  diag(\alpha_1,\ldots,\alpha_1;
\alpha_2,\ldots,\alpha_2;\ \ldots\ ; \alpha_p,\ldots,\alpha_p)
\ee
with $\alpha_i$ appearing $N_i$ times, $\sum_{i=1}^p N_i = N$.
In fact, there is no need to keep
these $N_i$ non-negative integers: in final expressions
they can be replaced by any complex numbers. Moreover,
$N_i$ can depend on $T_k$ (i.e. on the shape of $V(M)$).

Now, using at the intermediate stage the eigenvalue representation of matrix
integrals, one can rewrite \cite{David,KMT} the matrix integral (\ref{maint})
over $N\times N$ matrix $M$ as a $p$-matrix integral over $N_i\times N_i$
matrices $M_i$ (each obtained with the shift by $\alpha_i$: just changing
variables in the matrix integral (\ref{maint})), which is nothing but the
multi-cut solutions \cite{DV}
\be\label{Ni}
Z_V(t|M_0)\sim
\int \prod_{i=1}^p DM_i \exp\left(\sum_{i,k}\Tr t_k^{(i)}M^k \right)
\prod_{i<j}^p\alpha_{ij}^{2N_iN_j}\times
\ee
$$
\times
\exp \left(2\sum_{k,l=0}^\infty (-)^{k}\frac{(k+l-1)!}
{\alpha_{ij}^{k+l}k!l!}\Tr_iM_i^k\Tr_jM_j^l\right)
$$
The variables $t^{(i)}_k$ are given by the relation
\be
\sum_{k=0}^\infty t_k \left(\sum_{i=1}^p \Tr_i(\alpha_i + M_i)^k\right) =
\sum_{i=1}^p \left(\sum_{k=0}^\infty t_k^{(i)} \Tr_i M_i^k\right)
\label{ttalpha}
\ee
with arbitrary $N_i\times N_i$ matrices $M_i$.

Thus, we finally have $p-1$ basis functions that can be described by the
proper choices of the integration contours, and are associated with
different solutions to the saddle point equation.

\subsection{Resolvent and loop equation}

Another form of constraints (\ref{Vir2}) is produced by
rewriting the infinite set through the
unique generating function of all single trace correlators
\be
\rho^{(1)}(z|t)\equiv
\hat\nabla (z) {\cal F},\ \ \ \ \hat\nabla (z)\equiv\sum_{k\ge 0}
{1\over z^{k+1}}{\d\over\d t_k},\ \ {\cal F}\equiv g^2\log Z_V
\ee
Introducing the notation $v(z)$ for $\sum_k t_kz^k$,
one obtains {\it the loop equation} \cite{loop}
\be\label{loop}
\left[V'(z)\rho^{(1)}(z|t)\right]_-=
V'(z)\rho^{(1)}(z|t)-\left[V'(z)\rho^{(1)}(z|t)\right]_+=
\ee
$$
\left(\rho^{(1)}(z|t)\right)^2
+\left[v'(z)\rho^{(1)}(z|t)\right]_-+ g^2\hat\nabla(z)\rho^{(1)}(z|t)
$$
In order to consider (connected) multi-trace correlators, one needs
to introduce higher generating functions (also named loop mean, resolvent etc)
\be\label{corrho}
\rho^{(m)}(z_1,...,z_m|t)\equiv\hat\nabla (z_1)...\hat\nabla (z_m){\cal F}
\ee
Note that $G(z)$ introduced in (\ref{res}) is equal to $Z(T|t)\rho^{(1)}(z|t)$.
Moreover,
taken at all $t_k=0$, (\ref{loop}) reduces to (\ref{resdif}). However,
the quantity $\rho^{(1)}(z|0)$ generates only correlators of
single-trace operators (moments), which is not enough in the matrix case.
Therefore, in this case one needs to know the whole quantity
$\rho^{(1)}(z|t)$. It can be, however, expressed through
$\rho^{(m)}(z_1,...,z_m|0)$,
\be
\rho^{(1)}(z|t) = \sum_{m\geq 0} \frac{1}{m!}
\oint \ldots \oint v(z_1)\ldots v(z_m) \rho^{(m+1)}(z,z_1,\ldots,z_m|0)
\ee

\section{Genus zero solution}

One can solve the loop equations (\ref{loop}) recurrently (see, e.g.,
\cite{AMM1}) expanding them intosum of $\rho^{(m)}(z_1,...,z_m|0)$ and these
latter into series in $g$. This gives one the double recurrent relation (in
$m$ and order of $g$). Note that the solution of these recurrent relations
in the leading order can be immediately obtained. Indeed, omitting the last term
from (\ref{loop}), one obtains that
\be
\rho^{(0|1)}(z) = \frac{V'(z) - y(z)}{2}
\label{sdenW}
\ee
(where the first superscript 0 refers to the leading in $g$ approximation) with
\be\label{curve}
y^2(z) = (V'(z))^2 - 4P_{p-1}(z)
\ee
where $P_{p-1}(z)\equiv \left[V'(z)\rho^{(1)}(z|0)\right]_+$ is a polynomial
of degree $p-1$. Coefficients of this polynomial depend on $T_k$'s and first
derivatives of the arbitrary function of $p-1$ variables, which parameterize
solutions to constraints (\ref{Vir2}) (=the loop equations).

Note that formula (\ref{curve}) gives a hyperelliptic Riemann surface of
genus $p-1$, since mapping (\ref{curve}) gives $p$ cuts on the complex plane
(this is the celebrated multi-cut solution we discussed in the
introduction). It turns out that, at least with some specific fixing of the
ambiguity in solutions of the loop equations, the (multi-)resolvents can be
constructed as differentials given on this Riemann surface,
\cite{Ey,CMMV2}. This specific fixing is given by the conditions
\be\label{c1}
\oint_{A_i}\rho^{(m+1)}(z,z_1,\ldots,z_m|0)dz=0
\ee
for all the (multi-)resolvents except $\rho^{(0|1)}(z)$,
\be\label{c2}
\oint_{A_i}\rho^{(0|1)}(z)=S_i
\ee
where $S_i$ are arbitrary constants not depending on $T_k$ and
$A_i$ are $A$-cycles on the Riemann surface (\ref{curve}). The constants
$S_i$
can be, in fact, associated (up to the factor $g$) with $N_i$ from (\ref{Ni}).
Moreover, the basis functions (\ref{Ni}) exactly lead to conditions
(\ref{c1})-(\ref{c2}). These conditions are also associated with
Seiberg-Witten-Whitham system
corresponding to the Riemann surface (\ref{curve}) \cite{CM}.

We do not go into further details here, just referring to \cite{AMM3} and
\cite{CMMV2} for the latest development and proper references. Let us
only note that conditions (\ref{c1})-(\ref{c2}) distinguish specific
solutions that survive while smooth changing the number $p$ of couplings $T_k$,
much similar to what we observed in s.2.3. Indeed, for any number $q$
of non-zero $T_k$'s, one may require the curve to be of the form
\be\label{dp}
y^2(z)=(V'(z))^2 - 4P_{q-1}(z)=H_{q-p}^2(z)R_{2p}(z)
\ee
where $H_{q-p}^2(z)$ and $R_{2p}(z)$ are polynomials. Therefore,
independently of $q$, one has the hyperelliptic curve
${\tilde y}^2(z)=R_{2p}(z)$ of genus $p-1$. Moreover, the freedom one has in
matrix model in this case is dictated by $p-1$ constants $S_i$, see the
details in \cite{CMMV2}. The crucial difference from the case considered in
s.2.3 is that, when one is dealing with the ordinary integral, one may calculate
higher moments differentiating the integral in lower couplings several
times, while, in the matrix case, it is not possible because of the traces.
Therefore, technically this is very convenient to add traces of matrices in
higher degrees directly into the matrix potential. One, however, has to be
careful not to
change the solution of the matrix model with this procedure. This is what
exactly achieved with preserving the curve ${\tilde y}(z)$, (\ref{dp}).

\section*{Acknowledgments}

I am grateful to A.Alexandrov, L.Chekhov, A.Marshakov, A.Morozov and A.Vasiliev
for numerous discussions and to V.Dolotin for reading the manuscript.
The work is partly supported by Federal Program of the Russian Ministry of
Industry, Science and Technology No 40.052.1.1.1112, by the RFBR grant
04-01-00646a and the grant of Support for the Scientific
Schools 96-15-9679.

\end{document}